\documentclass[aps,prl,twocolumn,showkeys,groupaddress,preprintnumbers,floatfix]{revtex4-1}

\usepackage{dynlearn}
\usepackage{subfig}

\graphicspath{{./img/}}
\usepackage{braket}
\usepackage{multirow}
\usepackage{cleveref}
\definecolor{mygray}{gray}{0.6}

\theoremstyle{plain}    

\begin{document}

\def\ourTitle{Non-Markovian Momentum Computing: Universal and Efficient}

\def\ourAbstract{All computation is physically embedded. Reflecting this, a growing body of
results embraces rate equations as the underlying mechanics of thermodynamic
computation and biological information processing. Strictly applying the
implied continuous-time Markov chains, however, excludes a universe of natural
computing. We show that expanding the toolset to continuous-time \emph{hidden}
Markov chains substantially removes the constraints. The general point is made
concrete by our analyzing two eminently-useful computations that are impossible
to describe with a set of rate equations over the memory states. We design and
analyze a thermodynamically-costless bit flip, providing a first counterexample
to rate-equation modeling. We generalize this to a costless Fredkin gate---a key operation in reversible computing that is computation universal. Going beyond rate-equation dynamics is not only possible, but necessary if stochastic thermodynamics is to become part of the paradigm for physical information processing.
}

\def\ourKeywords{rate equations, stochastic process, hidden Markov model, information
processing, logical circuits, ion channel, entropy production, reversibility
}

\hypersetup{
  pdfauthor={James P. Crutchfield},
  pdftitle={\ourTitle},
  pdfsubject={\ourAbstract},
  pdfkeywords={\ourKeywords},
  pdfproducer={},
  pdfcreator={}
}

\author{Kyle J. Ray}
\email{kylejray@gmail.com }
\affiliation{Complexity Sciences Center and Physics Department,
University of California at Davis, One Shields Avenue, Davis, CA 95616}

\author{Alexander B. Boyd}
\email{alecboy@gmail.com}
\affiliation{Complexity Institute, Nanyang Technological University, 3 Science  Drive 2, Singapore 117543}

\author{Gregory W. Wimsatt}
\email{gwwimsatt@ucdavis.edu }
\affiliation{Complexity Sciences Center and Physics Department,
University of California at Davis, One Shields Avenue, Davis, CA 95616}

\author{James P. Crutchfield}
\email{chaos@ucdavis.edu}
\affiliation{Complexity Sciences Center and Physics Department,
University of California at Davis, One Shields Avenue, Davis, CA 95616}

\bibliographystyle{unsrt}

\title{\ourTitle}

\begin{abstract}
\ourAbstract
\end{abstract}


\preprint{\arxiv{2007.XXXXX}}

\title{\ourTitle}
\date{\today}
\maketitle

\setstretch{1.0}
\newcommand{\kB}{k_\text{B}}

The burgeoning field of thermodynamic computing leverages recent progress in
nonequilibrium thermodynamics and information and computation theories
\cite{Saga12a, Seif12a, parr2015,hase2019, seif2019} to establish a new
physical paradigm for computation \cite{Cont99a}. Working upwards from
fundamental laws of physics, it promises to increase computational efficiency
and power and to reduce energy dissipation in a next generation of computers.
More broadly, a general framework rooted in thermodynamics, as thermodynamic
computing is, will provide the tools to understand the physics of computation
in all its many forms. However, recent efforts inadvertently and unnecessarily
limit the potential scope.  The following illustrates the breadth of that scope
by introducing non-Markovian momentum computing that is both computation
universal and thermodynamically efficient.

A computation over a time interval $t \in (0,\tau)$ is described by the
conditional input-output mapping between memory states $m(0), m(\tau ) \in
\mathcal{M}$: $\mathbf{p}_{m(0) \rightarrow m(\tau)} = \Pr \left[ m(\tau) |
m(0)\right]$. The mapping $\mathbf{p}$ characterizes the probability of the
final memory state given the initial memory state.

Attempts to establish a general framework for the required mappings in
thermodynamic computing have assumed that the memory state $m$ is a physical
state of a system---it obeys stochastic Markovian dynamics, with time-evolution
depending only on the system's current state \cite[and references
therein]{Stop19a,Owen19a}. More broadly, weak coupling to a thermal bath and
separation of time scales are often invoked to justify such Markovian
thermodynamic behavior \cite{Seif12a, Alic79a, Jarz17a, Deff11a, Stra16a}.

Separating the memory system's time scale from that of the heat bath serves to
eliminate any memory in the heat bath from the system's behavior, as the heat
bath is assumed to rapidly relax to a local equilibrium. As a result,
transitions among memory states are potentially stochastic, but
Markovian---they only depend on the current state \cite{Seif12a}. Thus, these
systems obey continuous-time Markov chains (CTMCs), which are equivalently
represented by rate equations \cite{Stop19a,Owen19a}. Heat bath interactions
allow for a broader suite of behaviors among memory states than purely
deterministic Hamiltonian evolution allows on its own. However, this framework
is still restrictive. For example, only input-output mappings whose
determinants are positive are allowed when memory-state dynamics are restricted
to obey CTMCs. And this, for better or worse, eliminates a wide range of
possible and common computations, including flipping a single bit of
information \footnote{Such restrictions appear, for example, in Ref.
\cite{Stop19a}'s delineating what are ``physically realizable'' computations.}.

And so, while CTMCs are a powerful framework for stochastic thermodynamics
\cite{Vand13a, Vand15a, Seif12a}, they fail to capture a broad swath of
physically-realizable computing. They neglect the possibility of physical
variables that carry hidden memory of the past beyond the immediate
computational memory configuration.

In a thermodynamic system consisting of a collection of particle positions and momenta $(\vec{x},\vec{p})$ the dynamics are Langevin: stochastic, but Markovian \cite{kubo2012statistical}. The underlying dynamic that governs the combination of the system's full microstate and the thermal bath together, though, is deterministic and Hamiltonian. The system's stochastic evolution results from coarse-graining over the bath degrees of freedom. Focusing on the system's Markov dynamic alone is a conventional modeling strategy, especially since one is typically uninterested in the bath's details.

Similarly, if system memory-states are determined only by particle position,
that choice of state coarse-grains away an additional component of the system
and bath: system momentum. However, \emph{unlike} the microstates of the ideal
thermal bath, system momentum may carry memory of past behavior and this
contributes to memory-state dynamics. Soberingly, analytical treatment of
partially-observed (and therefore non-Markov) systems is highly nontrivial
\cite{koyu19, seif2019, maes17, Stra16a, stra19a}. That said, removing the CTMC
restriction permits realizing a broader range of computations. Thus, despite
the additional analytical burden, investigating systems that operate in the
regime where hidden states carry computationally-useful memory is a topic of
current focus \cite{fisc18, bran18, stum17, feri16, seif2019, horo17}. 

We argue that the appropriate setting for thermodynamic computing is
continuous-time \emph{hidden} Markov chains (CTHMCs), in which hidden variables
may store computationally-relevant information. Recently, this was recognized
as sufficient for a broad class of input-output mappings $\mathbf{p}$ by
introducing ancillary hidden states which implement sequences of logical
operations that individually obey CTMC dynamics \cite{Owen19a, bech2015}.
However, CTHMCs implement more general computations still.

The following first implements a thermodynamically-costless bit flip---a simple
computation that is explicitly forbidden by CTMCs. It then generalizes this to a
costless Fredkin gate \cite{Fred82a}---a key component in reversible computing
that is also impossible to implement with CTMCs. The implementation of this
universal and reversible logic gate via CTHMCs demonstrates that non-Markov
dynamics are essential to thermodynamic computing.

\paragraph*{Bit Flip}
To execute a single bit flip over a time interval $t \in [0,\tau]$, the first
step is to store a bit of information. One candidate is a particle with a
single position dimension $x \in \mathbb{R}$ and corresponding momentum $p \in
\mathbb{R}$ with double-well potential energy landscape: $V^\text{DW}(x) \equiv
\alpha x^4-\beta x^2$, where both $\alpha$ and $\beta$ are positive and
determine the location of the potential minima $x^*= \pm \sqrt{ \beta / 2
\alpha}$. The parameters are set so that information is stored robustly at the
beginning of the computation interval ($t = 0$).

The particle's environment is a thermal bath at temperature $T$. As the height
of the potential energy barrier at $x=0$ rises relative to the bath energy
scale $k_B T$, the probability that the particle transitions between left
($x<0$) and right ($x\geq0$) decreases exponentially. In this way, if we assign
the left half of the space to memory state $0$ and the right half to memory
state $1$, the energy landscape is capable of metastably storing a bit $m \in
\{0,1\}$.

To execute a flip operation, we instantaneously reduce the coupling to the
thermal reservoir to zero such that it now follows dissipationless Hamiltonian
dynamics. Simultaneously, the potential energy landscape changes to a positive
quadratic well: $V^\text{DW}(x,t=0^+)= kx^2 / 2$. The resulting particle motion
is harmonic oscillation: $x(t)=x^*\cos \left(t\sqrt{k/m}+ \phi \right)$, where
$x^*$ is the maximum distance from the origin of the cycle and $\phi$ is the
phase difference from maximum distance at the time $t=0^+$.  If we maintain the
decoupled system in the quadratic potential energy landscape for half the
period of oscillation $t \in \left(0, \pi \sqrt{m} / \sqrt{k} \right)$, then
the particle's new position becomes:
$x \left(\frac{\pi \sqrt{m}}{\sqrt{k}} \right) =x^*\cos \left(\pi+ \phi \right)
= -x^*\cos \left(\phi \right)
= -x(0)$.
Thus, over the computation interval $\tau= \pi \sqrt{m} / \sqrt{k}$, the
position flipped sign so that the memory state has flipped as well:
$m(\tau)=1-m(0)$. Finally, we instantaneously return the potential energy
landscape to the original double well and recouple to the thermal bath.

The work cost comes from changes in the potential energy at $t=0$ and $t=
\tau$. However, since particle position simply flips sign between $t=0$ and
$t=\tau$ and the potential energy landscape is even in position, zero net work
must be generated during this time-symmetric protocol.

Not only does this computation go beyond what is physically allowable according
to rate-equation dynamics over the memory states, but the states only change
while the Hamiltonian control is fixed.  Thus, the computation is passive,
meaning that it fits the information-ratchet framework introduced by Ref.
\cite{Boyd15a}.

\paragraph*{Fredkin Gate}
The bit-flip implementation may seem obvious in its simplicity. Can
sophisticated and functional computing, in fact, be built from such simple
passive processes? We answer in the affirmative by showing that a similar
strategy implements the Fredkin gate, a reversible and universal logical
gate \cite{Fred82a}. This straightforwardly establishes that the CTHMC
framework for thermodynamic computing gives easy access to complex and
universal Turing computing.

The \emph{Fredkin gate} operates on three bits $\mathcal{M}=\{0,1\}^3$. That
is, we encode our physical system as three particle-position variables
$(x,y,z)$ that are each separated into negative and positive memory-state
regions as above. This splits the memory states into eight respective
octants: $(x<0,y<0,z<0)$ corresponds memory state $m=000$, $(x<0,y\geq0,z<0)$ to
$m=010$, and so on. The information-storing Hamiltonian is a straightforward
sum of bistable double-wells in each dimension:
\begin{align*}
V^{\text{store}}(x,y,z)&=V^\text{DW}(x)+V^\text{DW}(y)+V^\text{DW}(z) \\
  &=\alpha \left(x^4+y^4+z^4\right)-\beta \left(x^2+y^2+z^2 \right)
  ~.
\end{align*}
This provides metastable regions corresponding to each memory state $m_x m_y m_z
\in \{0,1\}^3$.

Within this framework, we consider physical transformations that implement the
Fredkin gate and do so robustly. The Fredkin gate is also known as the
\emph{controlled swap gate}, as it swaps inputs $m_y$ and $m_z$ only if the
control $m_x$ is set to $1$. In other words, the gate maps all inputs to
themselves, excluding $101$ and $110$ which swap with each other. The
implementation uses the same strategy of decoupling and adding a harmonic
potential over the time interval $t \in (0, \tau)$, then recoupling and
resetting the original information-storing Hamiltonian. The only difference is
that the harmonic potential driving the computation is now embedded in the
higher-dimensional space.

To execute the Fredkin gate, first note that the memory-state $x$-index must
always be fixed: $m_x(\tau)=m_x(0)$. Moreover, behavior in the $y-z$ plane
should only depend on $x$ up to whether it is positive or negative. Thus, we
first split the potential into two pieces: $V(x, y, z, t) = V^\text{DW}(x) +
V^{yz}(x, y, z, t)$. If $m_x(0)=0$ then $m_y$ and $m_z$ must also not change.
This suggests using the information-storing potential for this region of state
space: $V(x<0,y,z,t) = V^{\text{store}}(x,y,z)$, so that:
\begin{align*}
V^{yz}(x<0,y,z,t) = V^\text{DW}(y) + V^\text{DW}(z)
  ~.
\end{align*}
For $m_x=1$, however, we must nontrivially compute on $m_y$ and $m_z$:
\begin{align*}
V^{yz}(x\geq0, y, z, t \in (0, \tau)) = V^\text{comp}(y,z)
  ~.
\end{align*}
Here, $V^\text{comp}$ determines that part of the Hamiltonian which implements
the switch $101 \rightarrow 110$ and $110 \rightarrow 101$ and remains
unchanging over $t\in(0,\tau)$. Due to decoupling from the $x$-axis, particle
behavior in either the positive or negative $x$ regions can be considered as
being purely the result of two-dimensional dynamics.

\begin{figure}[t]
\centering
\includegraphics[width=\columnwidth]{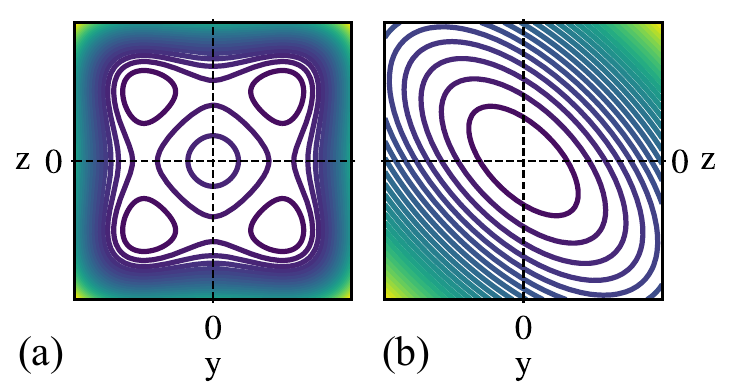}
\caption{Potential $V^{yz}$ in the $y-z$ plane in the (a) information-storing
($x<0$) and (b) controlled-swap ($x>0$) domains.
	}
\label{fig:V_comp_store}
\end{figure}

To swap $101$ and $110$, while keeping $111$ and $100$ fixed, consider a new
basis for the $yz$-space. Define new variables: $y' = (y-z)/\sqrt{2}$ and $z' =
(y+z)/\sqrt{2}$, such that the local equilibrium distributions for states $110$
and $101$ are centered around $z'=0$ and those for states $111$ and $100$ are
centered around $y'=0$. Thus, our goal is to swap the distributions in the
$y'$-coordinate while preserving their $z'$-coordinate. Given this, we split
the computation Hamiltonian again into independent components:
$V^\text{comp}(y,z) = V(y') + V(z')$. Flipping in the $y'$-coordinate employs
the same Hamiltonian as for the previous Bit Flip protocol: $V(y') = ky'^2 /
2$. As a result, when waiting half a period $\tau= \pi \sqrt{m} / \sqrt{k}$,
the $y'$ coordinate changes sign $y'(\tau)=-y'(0)$, as does its momentum. We
choose the $z'$ coordinate's potential to be quadratic as well, but with an
induced period of oscillation that is half as long: $V(z') = 2kz'^2$. $z'$ then
undergoes a full cycle after the duration $\tau = \pi \sqrt{m} / \sqrt{k}$,
returning to its original value $z'(\tau)=z(0)$, as does its momentum.

\begin{figure*}[t]
\centering
\includegraphics[width=2\columnwidth]{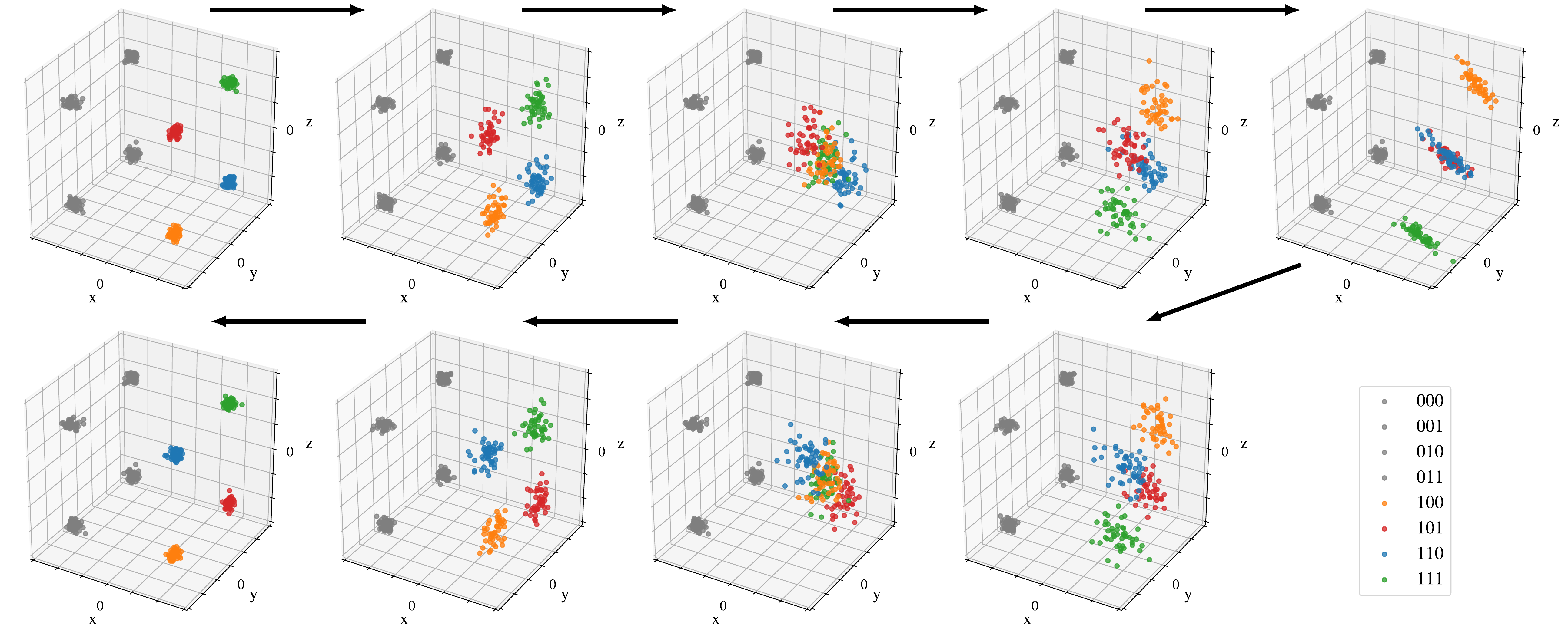}
\caption{Particle ensemble initialized in equilibrium with
	$V^\text{store}(x,y,z)$ undergoing the Fredkin gate protocol with zero
	coupling to the thermal reservoir. Each snapshot of the state evolution is
	separated by a time interval of $\tau/8$, with the black arrows indicating
	forward time. Color encodes in which informational state each trial begins.
	The $101$ (red) and $110$ (blue) states only oscillate by a quarter period
	in the time ($\tau/2$) it takes the $100$ (yellow) and $111$ (green) states
	to oscillate by a half cycle. As the $100$ and $111$ trials return to their
	initial positions, the $110$ and $101$ states approach their final
	positions: a half cycle from where they started (right). The states have
	been swapped.
	\href{http://csc.ucdavis.edu/~cmg/compmech/pubs/cbdb.htm}{Animations
	available online}.
	}
\label{fig:zero_coupling}
\end{figure*}

The resulting full Hamiltonian over the control interval operates piecewise. 
Figure \ref{fig:V_comp_store} shows the potential in the $x<0$ and $x>0$ regions
during the computation interval:
\begin{align}
V(x,y',z',t)
  &= V^{DW}(x) + V^{yz}(x, y', z', t)&\\
  &= \begin{cases}
  V^{\text{store}} & \text{if }x<0 \\
  V^{DW}(x) + \frac{ky'^2}{2} + 2kz'^2 & \text{if } x\geq 0
  \end{cases}
  ,
\label{eq:fredkin_potential}
\end{align}
for $t\in (0,\tau)$.
Translating back to the original coordinates $y= (y'+z')/\sqrt{2}$ and $z=
(z'-y')/\sqrt{2}$, we find that for $x\geq0$, this passive Hamiltonian
transforms the particle's state by swapping $y$ and $z$:
\begin{align*}
(y(\tau),z(\tau))
  & =\left(
  \frac{y'(\tau)+z'(\tau)}{\sqrt{2}}, \frac{-y'(\tau)+z'(\tau)}{\sqrt{2}}
  \right) \\
  & =\left(
  \frac{-y'(0)+z'(0)}{\sqrt{2}}, \frac{y'(0)+z'(0)}{\sqrt{2}}
  \right) \\
  & = (z(0),y(0))
  ~,
\end{align*}
while it holds the other four quadrants where $m_x=0$ in their respective potential minima.
Thus, the transformation swapped $y$ and $z$ only when $m_x=1$, implementing the
Fredkin gate.

For a particular trajectory $(x,y,z)(t)$, the work invested only comes from the
initial and final instantaneous changes in the energy landscape:
\begin{align*}
W&=V(x(0),y(0),z(0),0^+)-V(x(0),y(0),z(0),0) \\
  & ~~+V(x(\tau),y(\tau),z(\tau),\tau)-V(x(\tau),y(\tau),z(\tau),\tau^-)
  ~.
\end{align*}
Recall the restriction that $x(t)$ is exponentially unlikely to change sign,
because the energy barrier between states is much higher than the vast majority
of thermal fluctuations can access. Thus, we assume that paths maintain a
single sign for $x(t)$. If $x(t)$ is negative, then there is no instantaneous
change, as the system is held in the same double-well potential, so $W=0$.
That said, if $x(t)$ is positive, then the work invested also vanishes.
 
The $x$ subspace of the potential decouples from the $y-z$ subspace and remains
constant. Thus, there are no work contributions from the $x$-dependent terms.
Additionally, the $y-z$ subspace potential is symmetric with respect to
exchange of the $y$ and $z$ coordinates. So, the energy differences above will
vanish for the $y$ and $z$ dependent terms as well. (Recall that the action of
the potential over our interval is to swap the $y$ and $z$ coordinates so that
$(y(\tau),z(\tau)) = ((z(0), y(0))$.) And so, the average work production is
nearly zero---only the exponentially suppressed barrier crossing events can
contribute to non-zero work values.

Figure \ref{fig:zero_coupling} demonstrates the evolution of the phase space on
an ensemble of initial conditions drawn from the equilibrium distribution of
$V^{\text{store}}(x,y,z)$. As shown by the particle coloring, those that start
in $110$ and $101$ swap while all others are fixed. Moreover, none of the
particles' $x$-coordinate change informationally---confirming the effectiveness
of the overall transformation.

\newcommand\ntrials{20000 }
\newcommand\nsteps{6000 }
\newcommand\ethresh{1000}
\newcommand\aparam{2}
\newcommand\bparam{16}
\newcommand\kparam{1}

\paragraph*{Langevin Simulation}
The preceding stipulated that the logical system be isolated from its thermal
environment during the swap. However, the impact of adding thermal coupling is
minimal. To demonstrate this, we investigated how robust the operation is to
thermal coupling by using \emph{underdamped} Langevin dynamics. A simulation was
carried out by initializing \ntrials particles in equilibrium with a thermal
reservoir under the information-storing potential $V^\text{store}(x,y,z)$.
Next, as described above, we exert work on the system by turning on the
computational potential $V^\text{comp}$ in the region $x>0$. However, rather
than reducing the thermal coupling to $\lambda=0$, we drop the coupling
coefficient to a nonzero value in the weak coupling regime. This coupling value
and potential are held fixed for time $\tau= \pi \sqrt{m/k}$. (The appendix
provides additional detail.)

\begin{figure}[t]
\centering
\includegraphics[width=\columnwidth]{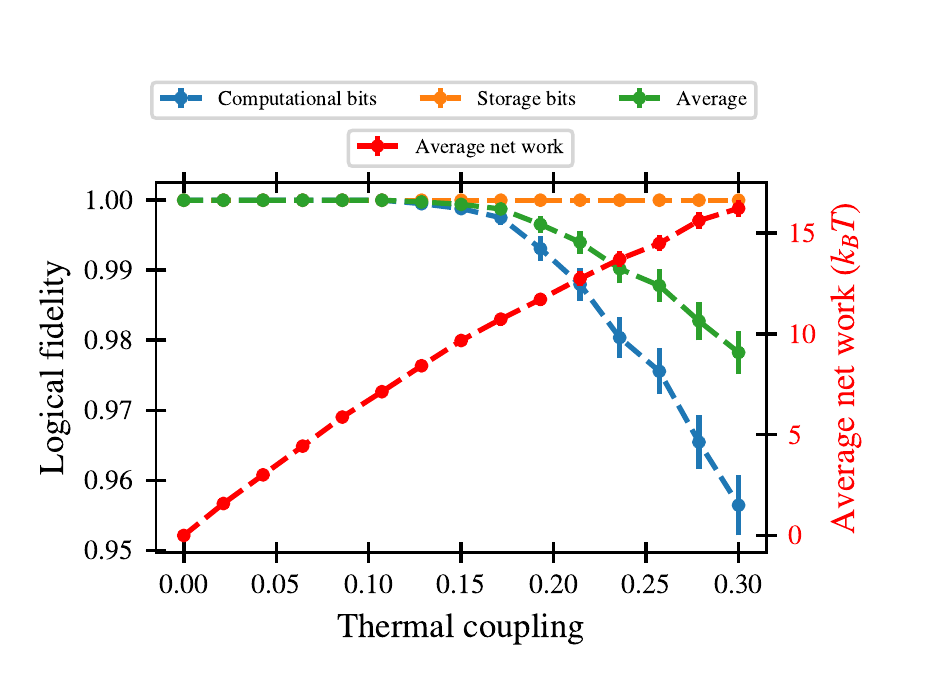}
\caption{Logical fidelity ($\text{successful trials}/\text{total trials}$) in
	the low-coupling Fredkin gate and the average net work required to
	implement it for different values of the thermal coupling constant
	$\lambda$. \emph{Computational bits} refers to states that fall in the
	region $x>0$, where the computational potential is in effect.
	}
\label{fig:coupling_scaling}
\end{figure}

The particles experience thermal fluctuations as the weak coupling to the bath
perturbs their trajectories from the otherwise expected harmonic motion. The
work gained from shutting off the potential will not generally be the same as
the work invested to turn it on (as in the idealized case of zero thermal
coupling). In fact, the Second Law guarantees that, generally, positive work is
invested for such cyclical transformations, because the net change in
equilibrium free energy is zero. Nevertheless, one expects the behavior to
approximate the desired Fredkin-gate dynamics if the coupling is sufficiently
weak. While the energetic cost of implementing the gate does not remain zero as
the coupling approaches zero, Fig. \ref{fig:coupling_scaling} shows that the
logical fidelity approaches unity. And, it does so with \emph{zero slope},
revealing that this Fredkin gate implementation is robust even in the presence
of thermal fluctuations. Thus, we see that the Fredkin gate (CTHMC) dynamics
do not rely on removing the thermal reservoir.

As expected and shown in Fig. \ref{fig:coupling_scaling}, the work invested
approaches zero with decreasing coupling. However, as the coupling to the
thermal reservoir increases, the average work required to compute increases
to multiples of $k_B T$. This cost is much more than predicted by
the microscopic detailed-balance dynamics that underlie the Langevin
simulation. This suggests the existence of a lower bound on entropy
production---one that accounts for the course-graining, as predicted in
Ref. \cite{Riec19b}.

\paragraph*{Conclusion}

Rate equation dynamics is certainly a venerable and powerful framework, central
to reaction kinetics in chemistry \cite{Laid84a, Stei99a} and key to the master
equations of applied statistical mechanics \cite{Vand13a, Vand15a, Seif12a}. In
fact, perhaps due to the remarkable successes of continuous-time Markov chain
predictions of many thermodynamic behaviors, it might seem natural to claim
that in order to be ``physically realizable'', thermodynamic computing and
biological information processing can \emph{only} be described and analyzed as
rate-equation dynamics \cite{Stop19a}. 

However, we demonstrated this framework cannot form a complete basis for
thermodynamic computing. Moreover, its strict application levies a penalty that
precludes engineering and analyzing Maxwellian information ratchets, which are
the physical equivalent of Turing machines
\cite{Boyd15a,Boyd16e,Jurg20a,Boyd17a}. The limits are especially draconian,
since efficient time-symmetrically controlled general computations consist of
involutions \cite{Riec19b}---operations that are composed of bit swaps and
identity maps in positional memory (or any memory that is even under time
reversal).

As a constructive alternative, we proposed employing continuous-time
hidden Markov chains to realize \emph{non-Markovian momentum computing}.
We demonstrated it provides a more complete framework, using two explicit
examples that are forbidden if one is restricted to rate equations to describe
the evolution between memory states \cite{Stop19a}. More and helpfully, we
introduced explicit mechanisms for implementing both with zero work, proving
that they are most certainly ``physically realizable''.

Not only are hidden Markov chains more general, but their added generality is
critical in many circumstances. The fact that the Fredkin gate can be executed
robustly, even when thermal fluctuations perturb the particle trajectories,
suggests that this implementation will have practical use for reversible
universal computing. The robustness of the gate to fluctuations separates it
from other implementations of reversible computing---such as, ballistic
computing with billiards---that are dynamically unstable \cite{Fred82a}.

We did, however, fully acknowledge the increased analytical complexity posed by
CTHMC dynamics. Fortunately, the requisite tools have been developed that
render the behaviors analytically tractable and in closed form
\cite{Riec16b,Riec16a}. In short, there is little impediment to reaching the
full generality of thermodynamic computing with CTHMCs.

Given that convincing physically-realizable implementations of the bit flip and
and Fredkin gate \cite{Fred82a, Milb89a, Wenz14a} have been known for some
time, one can only conclude that computing devices must be able to operate
beyond the restrictions imposed by rate-equation dynamics. The examples
presented here were intentionally couched in the thermodynamics of information
to help bridge an apparent gap in understanding general computing. Most
specifically, the conception of memory must be modified, from being the
realization of a microscopic  physical state to being a mesoscopic
coarse-graining, to fully realize the power and breadth of physical
computations.  

\paragraph*{Acknowledgments.}
\label{sec:acknowledgments}

We thank Adam Kunesh, Thomas Ouldridge, and Mikhael Semaan for helpful
discussions. JPC thanks the Santa Fe Institute and he and the other authors
thank the Telluride Science Research Center for their hospitality during
visits. This material is based upon work supported by, or in part by, FQXi
Grant number FQXi-RFP-IPW-1902, the Templeton World Charity Foundation Power of
Information fellowship TWCF0337, the U.S. Army Research Laboratory and the U.
S. Army Research Office under contract W911NF-13-1-0390 and grant
W911NF-18-1-0028, and via Intel Corporation support of CSC as an Intel Parallel
Computing Center.


\begin{thebibliography}{10}

\bibitem{Saga12a}
T.~Sagawa.
\newblock Thermodynamics of information processing in small systems.
\newblock {\em Prog. Theo. Phys.}, 127(1), 2012.

\bibitem{Seif12a}
U.~Seifert.
\newblock Stochastic thermodynamics, fluctuation theorems and molecular
  machines.
\newblock {\em Rep. Prog. Physics}, 75(12), 2012.

\bibitem{parr2015}
J.~M.~R. Parrondo, J.~M. Horowitz, and T.~Sagawa.
\newblock Thermodynamics of information.
\newblock {\em Nature Physics}, 11(2):131--139, 2015.

\bibitem{hase2019}
Y.~Hasegawa and T.~Van~Vu.
\newblock Fluctuation theorem uncertainty relation.
\newblock {\em Phys. Rev. Lett.}, 123(11):110602, 2019.

\bibitem{seif2019}
U.~Seifert.
\newblock From stochastic thermodynamics to thermodynamic inference.
\newblock {\em Ann. Rev. Cond. Mat. Physics}, 10:171--192, 2019.

\bibitem{Cont99a}
T.~Conte et~al.
\newblock Thermodynamic computing.
\newblock {\em arxiv:1911.01968}.

\bibitem{Stop19a}
E.~Stopnitzky, S.~Still, T.~E. Ouldridge, and L.~Altenberg.
\newblock Physical limitations of work extraction from temporal correlations.
\newblock {\em Phys. Rev. E}, 99:042115, 2019.

\bibitem{Owen19a}
J.~A. Owen, A.~Kolchinsky, and D.~H. Wolpert.
\newblock Number of hidden states needed to physically implement a given
  conditional distribution.
\newblock {\em New J. Physics}, 21(1):013022, January 2019.

\bibitem{Alic79a}
R.~Alicki.
\newblock The quantum open system as a model of the heat engine.
\newblock {\em J. Phys. A}, 12(5), 1979.

\bibitem{Jarz17a}
C.~Jarzynski.
\newblock Stochastic and macroscopic thermodynamics of strongly coupled
  systems.
\newblock {\em Phys. Rev. X}, 7(011008), 2017.

\bibitem{Deff11a}
S.~Deffner and E.~Lutz.
\newblock Nonequilibrium entropy production for open quantum systems.
\newblock {\em Phys. Rev. Lett.}, 107(140404), 2011.

\bibitem{Stra16a}
P.~Strasberg, G.~Schaller, N.~Lambert, and T.~Brandes.
\newblock Nonequilibrium thermodynamics in the strong coupling and
  non-{Markovian} regime based on a reaction coordinate mapping.
\newblock {\em New J. Physics}, 18(073007), 2016.

\bibitem{Note1}
Such restrictions appear, for example, in Ref. \cite {Stop19a}'s delineating
  what are ``physically realizable'' computations.

\bibitem{Vand13a}
C.~van~den Broeck.
\newblock Stochastic thermodynamics: A brief introduction.
\newblock {\em Phys. Complex Colloids}, 2013.

\bibitem{Vand15a}
C.~van~den Broeck and M.~Esposito.
\newblock Ensemble and trajectory thermodynamics: A brief introduction.
\newblock {\em Phyisca A}, 418:6--16, 2015.

\bibitem{kubo2012statistical}
R.~Kubo, M.~Toda, and N.~Hashitsume.
\newblock {\em Statistical physics {II}: Nonequilibrium statistical mechanics},
  volume~31.
\newblock Springer Science \& Business Media, 2012.

\bibitem{koyu19}
T.~Koyuk and U.~Seifert.
\newblock Operationally accessible bounds on fluctuations and entropy
  production in periodically driven systems.
\newblock {\em Phys. Rev. Lett.}, 122(23):230601, 2019.

\bibitem{maes17}
C.~Maes.
\newblock Frenetic bounds on the entropy production.
\newblock {\em Phys. Rev. Lett.}, 119(16):160601, 2017.

\bibitem{stra19a}
P.~Strasberg and M.~Esposito.
\newblock {Non-Markovianity} and negative entropy production.
\newblock {\em Phys. Rev. E}, 99(1)(012120), 2019.

\bibitem{fisc18}
L.~P. Fischer, P.~Pietzonka, and U.~Seifert.
\newblock Large deviation function for a driven underdamped particle in a
  periodic potential.
\newblock {\em Phys. Rev. E}, 97(2):022143, 2018.

\bibitem{bran18}
K.~Brandner, T.~Hanazato, and K.~Saito.
\newblock Thermodynamic bounds on precision in ballistic multiterminal
  transport.
\newblock {\em Phys. Rev. Lett.}, 120(9):090601, 2018.

\bibitem{stum17}
P.~S. Stumpf, R.~C.~G. Smith, M.~Lenz, A.~Schuppert, F-J. M{\"u}ller,
  A.~Babtie, T.~E. Chan, M.~P.~H. Stumpf, C.~P. Please, S.~D. Howison, et~al.
\newblock Stem cell differentiation as a {non-Markov} stochastic process.
\newblock {\em Cell Systems}, 5(3):268--282, 2017.

\bibitem{feri16}
L.~Ferialdi.
\newblock Exact closed master equation for {Gaussian Non-Markovian} dynamics.
\newblock {\em Phys. Rev. Lett.}, 116:120402, Mar 2016.

\bibitem{horo17}
J.~M. Horowitz and T.~R. Gingrich.
\newblock Proof of the finite-time thermodynamic uncertainty relation for
  steady-state currents.
\newblock {\em Phys. Rev. E}, 96(2):020103, 2017.

\bibitem{bech2015}
J.~Bechhoefer.
\newblock Hidden {Markov} models for stochastic thermodynamics.
\newblock {\em New J. Physics}, 17(7):075003, 2015.

\bibitem{Fred82a}
E.~Fredkin and T.~Toffoli.
\newblock Conservative logic.
\newblock {\em Intl. J. Theo. Phys.}, 21(3-4):219--253, 1982.

\bibitem{Boyd15a}
A.~B. Boyd, D.~Mandal, and J.~P. Crutchfield.
\newblock Identifying functional thermodynamics in autonomous {Maxwellian}
  ratchets.
\newblock {\em New J. Physics}, 18:023049, 2016.

\bibitem{Riec19b}
P.~M. Riechers, A.~B. Boyd, G.~W. Wimsatt, and J.~P. Crutchfield.
\newblock Balancing error and dissipation in computing.
\newblock {\em Phys. Rev. Res.}, 2(3):033524, 2020.
\newblock arXiv:1909.06650.

\bibitem{Laid84a}
K.~J. Laidler.
\newblock The development of the {Arrhenius} equation.
\newblock {\em J. Chem. Edu.}, 61.6(494), 1984.

\bibitem{Stei99a}
J.~I. Steinfeld, J.~S. Francisco, and W.~L. Hase.
\newblock {\em Chemical kinetics and dynamics}.
\newblock Prentice Hall, 1999.

\bibitem{Boyd16e}
A.~B. Boyd, D.~Mandal, P.~M. Riechers, and J.~P. Crutchfield.
\newblock Transient dissipation and structural costs of physical information
  transduction.
\newblock {\em Phys. Rev. Lett.}, 118:220602, 2017.

\bibitem{Jurg20a}
A.~Jurgens and J.~P. Crutchfield.
\newblock Functional thermodynamics of maxwellian ratchets: Constructing and
  deconstructing patterns, randomizing and derandomizing behaviors.
\newblock {\em Phys. Rev. Res.}, 2(3):033334, 2020.

\bibitem{Boyd17a}
A.~B. Boyd, D.~Mandal, and J.~P. Crutchfield.
\newblock Thermodynamics of modularity: Structural costs beyond the {Landauer}
  bound.
\newblock {\em Phys. Rev. X}, 8(3):031036, 2018.

\bibitem{Riec16b}
P.~M. Riechers and J.~P. Crutchfield.
\newblock Fluctuations when driving between nonequilibrium steady states.
\newblock {\em J. Stat. Phys.}, 168(4):873--918, 2017.

\bibitem{Riec16a}
P.~M. Riechers and J.~P. Crutchfield.
\newblock Beyond the spectral theorem: Decomposing arbitrary functions of
  nondiagonalizable operators.
\newblock {\em AIP Advances}, 8:065305, 2018.

\bibitem{Milb89a}
G.~J. Milburn.
\newblock Quantum optical {Fredkin} gate.
\newblock {\em Phys. Rev. Lett.}, 62(18), 1989.

\bibitem{Wenz14a}
J.-S. Wenzler, T.~Dunn, T.~Toffoli, and P.~Mohanty.
\newblock A nanomechanical {Fredkin} gate.
\newblock {\em Nano Lett.}, 14(1):89--93, 2014.

\end{thebibliography}

\onecolumngrid
\clearpage
\begin{center}
\large{Supplementary Materials}\\
\vspace{0.1in}
\emph{\ourTitle}\\
\vspace{0.1in}
{\small
Kyle J. Ray,
Alexander B. Boyd,
Gregory W. Wimsatt, and
James P. Crutchfield
}
\end{center}

\appendix

\section{Langevin Simulations}
\label{app:sim}

Simulations were carried out using Langevin equations of motion:
\begin{align*}
  dx &= v dt \\
  dv &= -\lambda v dt - \partial_x U(x,t) dt + \sqrt{2\kB T \lambda}\, r(t) \sqrt{dt}
  ~,
\end{align*}
where $r(t)$ is a memoryless Gaussian random variable with zero mean and unit
variance. Since we track behavior when sweeping the thermal coupling parameter
$\lambda$ only, it is convenient to consider a particle with unit mass ($m=1$)
and set $\kB T =1$. This yields a very simple dynamic that is readily
interpreted:
\begin{align*}
  dx &= v dt \\
  dv &= -\lambda v dt -  \partial_x U(x,t) dt + \sqrt{2 \lambda}\, r(t) \sqrt{dt}
  ~.
\end{align*}
The parameter $\lambda$ (the \emph{thermal coupling coefficient}) controls the
damping force the particle experiences from the thermal bath when it has
unit velocity. It is commonly called the damping coefficient or the inverse
mobility.

The simulation employed the fourth-order Runge-Kutta method for the
deterministic portion and Euler's method for stochastic portion of the
integration. (Python NumPy's Gaussian number generator was used to generate the
memoryless Gaussian variable r(t).)

Figure \ref{fig:coupling_scaling}'s plot displays $3\sigma$ error bars, but the
errors are generally small enough that they do not show up appreciably.
Statistical errors were estimated using standard procedures for sample means
and proportions.

Figure \ref{fig:coupling_scaling} was generated from simulation using the
following procedure. First, an ensemble of \ntrials trials were chosen from an
approximate equilibrium distribution of $V^{\text{store}}(x,y,z)$ with $\alpha
= \aparam, \beta=\bparam$, using the Monte Carlo algorithm. Second, this
ensemble was thermalized while coupled to a bath ($\lambda=1$) until the
ensemble energy changed by no more than $1$ part in \ethresh\, over a unit time
interval. Third, this ensemble was then used as the start state for the Fredkin
gate operation. We then dropped $\lambda$ down to a low coupling value and
exposed the unit mass particles to the potential in Eq.
(\ref{eq:fredkin_potential}) with $\alpha = \aparam$, $\beta=\bparam$, and
$k=\kparam$. Fourth, at this point we measured the work required to change the
potential across our ensemble. Fifth, the potential was then held fixed for a
time $\tau = \pi \sqrt{k/m} = \pi $ using an integration step $dt \approx
0.0005$. Finally, immediately following the computation interval, we measured
the second work contribution---the work that would be harvested by dropping the
potential back to $V^{\text{store}}$. The average net work is the ensemble
average difference between the work invested when raising the potential and the
work harvested when lowering it.

Figure \ref{fig:zero_coupling} was generated by starting the particles in the
equilibrium distribution described above, and running the simulation with
$\lambda=0$, to simulate dissipationless oscillatory dynamics.  The plot shows
a sample of $200$ trials, rather than the full $20000$, for clarity. 

\end{document}